\documentclass[aps,pra,twocolumn,groupedaddress]{revtex4-1}

\usepackage{graphicx}
\usepackage{color}
\usepackage{bm}
\usepackage{amsmath}
\usepackage{braket}

\usepackage[utf8]{inputenc} 
\newcommand{\sech}{\mathrm{sech} \,}

\newcommand{\eref}[1]{(\ref{#1})}

\begin{document}

\title{Limits on the deterministic creation of pure single-photon states using parametric down-conversion}

\author{Andreas Christ}
\email{andreas.christ@uni-paderborn.de}
\author{Christine Silberhorn}
\affiliation{Applied Physics, University of Paderborn, Warburger Stra{\ss}e 100, D-33098 Paderborn}
\affiliation{Max Planck Institute for the Science of Light, 
G\"unther-Scharowsky-Str. 1/Building 24, D-91058 Erlangen, Germany}


\begin{abstract}
Parametric down-conversion (PDC) is one of the most widely used methods to create pure single-photon states for quantum information applications. However little attention has been paid to higher-order photon components in the PDC process, yet these ultimately limit the prospects of generating single photons of high quality. In this paper we investigate the impact of higher-order photon components \textit{and} multiple frequency modes on the heralding rates and single-photon fidelities. This enables us to determine the limits of PDC sources for single-photon generation. Our results show that a perfectly single-mode PDC source in conjunction with a photon-number-resolving detector is ultimately capable of creating single-photon Fock states with unit fidelity and a maximal state creation probability of 25\%. Hence, an array of 17 switched sources is required to build a deterministic (\(>99\%\) emission probability) pure single-photon source. 
\end{abstract}


\maketitle

\section{Introduction\label{sec:introduction}}
\textit{Pure} single-photon states are an essential ingredient for quantum information technologies such as quantum communication \cite{gisin_quantum_2007}, quantum enhanced measurements \cite{giovannetti_quantum-enhanced_2004} and quantum computing \cite{walmsley_toward_2005}. In the past decades various sources have been investigated to produce the required \textit{pure} single-photon states including semiconductor quantum dots \cite{michler_quantum_2000, santori_indistinguishable_2002}, trapped atoms \cite{kuhn_deterministic_2002, beugnon_quantum_2006}, trapped ions \cite{maunz_quantum_2007, barros_deterministic_2009} and four-wave-mixing processes \cite{rarity_photonic_2005, chen_fiber-based_2006,fan_broadband_2007,fulconis_nonclassical_2007, smith_photon_2009, ling_mode_2009,soller_bridging_2010, soller_high-performance_2011}. To date, however, the most widely used sources for the creation of single photons are still based on parametric down-conversion (PDC) \cite{hong_measurement_1987, castelletto_optimizing_2006, pittman_heralding_2005, uren_efficient_2004,lvovsky_quantum_2001}  where substantial efforts haven been made over the past several years to engineer photon-pairs with single-mode characteristics \cite{mosley_heralded_2008, eckstein_highly_2011, gerrits_generation_2011, evans_bright_2010, poh_eliminating_2009}. 

PDC sources feature many advantages: The setups are compact, cost effective, robust, operate at room temperature, and can be integrated in optical circuits. However, they also possess some inherent drawbacks: First, the photon heralding is a statistical process and, hence, PDC always only approximates a deterministic single-photon source. Second, multi-photon-pair emission \cite{sekatski_detector_2011, huang_photon-counting_1989, osullivan_conditional_2008, rohde_photon_2007, rohde_improving_2007, huang_optimized_2011, wasilewski_statistics_2008, achilles_direct_2006, mauerer_how_2009, broome_reducing_2011} limits the heralding rates and the fidelity of the generated single-photon states. Finally, the spectral properties of the source may lead to a heralding of single photons in a mixture of frequency modes, diminishing the purity of the heralded state.

In this paper we investigate the trade-off between heralding rates and the fidelity of the heralded states using PDC processes extending the work presented in Refs. \cite{virally_limits_2010} and \cite{osullivan_conditional_2008}. We consider both binary avalanche photodiode detectors, as currently employed in laboratories, but also extend our analysis to incorporate the rapidly growing field of photon-number-resolved detection  \cite{fitch_photon-number_2003,divochiy_superconducting_2008,kardynal_avalanche-photodiode-based_2008, miccaronuda_high-efficiency_2008, fujiwara_multiphoton_2005}. Our results quantify the definitive limits of parametric down-conversion sources to create pure single-photon states and show how well they are able to approximate deterministic behavior.

\section{PDC state generation}
\begin{figure}[htp]
    \begin{center}
        \includegraphics[width=0.9\linewidth]{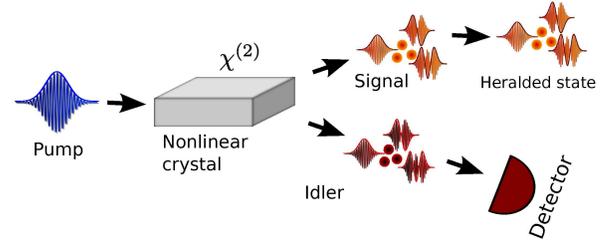}
    \end{center}
    \caption{Schematic of the PDC process used to herald single-photon states: An incoming pump pulse decays inside a nonlinear medium into two beams labeled signal and idler which feature a perfect photon-number correlation. The idler beam is subsequently detected to herald the presence of the signal state.} 
    \label{fig:single_photon_generation_process}
\end{figure}

Fig. \ref{fig:single_photon_generation_process} sketches the process of parametric down-conversion using a pulsed laser system. The incoming pump interacts with the crystal material featuring a \(\chi^{(2)}\) nonlinearity creating two down-converted beams usually labeled signal and idler. These two beams exhibit perfect correlation in photon number, which means that during the interaction a certain number of photon pairs is generated depending on the efficiency of the PDC.

The process is, in the interaction picture, described by the following Hamiltonian 
\begin{eqnarray}
    \nonumber
    \hat{H}_{PDC} \propto \chi^{(2)} \int \mathrm d^3 r \, \hat{E}^{(-)}_p(\vec{r},t) \hat{E}^{(+)}_s(\vec{r},t)\hat{E}^{(+)}_i(\vec{r},t) + h.c.
    \label{eq:pdc_hamiltonian}
\end{eqnarray}
where we consider both the spatial and spectral-temporal degree of freedom. Solving this Hamiltonian \cite{christ_probing_2011, christ_spatial_2009}, assuming a non depleted classical pump laser to drive the down-conversion process, we obtain the following PDC state:
\begin{align}
    \nonumber
    \ket{\psi}_{PDC} =  \exp\left[-\frac{\imath}{\hbar} \left(B \sum_{k,l} \iint \mathrm d\omega_s \, d \omega_i \, f_{k,l}(\omega_s, \omega_i) \right. \right. \\ \times \left. \left .\hat{a}_k^{(s)\dagger}(\omega_s) \hat{a}_l^{(i)\dagger}(\omega_i) + h.c. \right)\right] \ket{0}.
    \label{eq:pdc_state_multimode}
\end{align}
The operators \(\hat{a}_k^{(s)\dagger}(\omega_s)\) and \(\hat{a}_l^{(i)\dagger}(\omega_i)\) create photons with spatial mode numbers \(k\) and \(l\) and frequencies \(\omega_s\) and \(\omega_i\) into the signal and idler beam, respectively. The exact form of the output state is given by the function \(f_{k,l}(\omega_s, \omega_i)\) describing its spectral and spatial structure depending on the applied pump beam and nonlinear optical material \footnote{In the high-gain regime time-ordering effects have to be considered in the derivation of Eq. \eref{eq:pdc_state_multimode} \cite{wasilewski_pulsed_2006}.}. 

The spectral and spatial degrees of freedom are the first obstacle for the heralding of pure single-photon states. Since the photons are emitted into a multitude of spatial and spectral modes the detection of the idler beam to herald the presence of the signal results in a projection of the signal state into a mixture of spatial and spectral modes. Hence, the heralded signal does \textit{not} form a \textit{pure} single-photon state.

The easiest solution to cope with this problem is to apply heavy spectral and spatial filtering in the heralding arm \cite{huang_optimized_2011, smirr_optimal_2011, branczyk_optimized_2010}.  This will eliminate all distinguishing features and project the heralded signal into a spectrally as well as spatially pure state. However, one should be aware of the fact that the applied filter absorbs the main part of the generated idler photons and, hence, leads to significantly lower heralding rates and, furthermore, increases the higher-order photon components in the signal arm, negatively affecting the state fidelity in the photon number degree of freedom.

A more elegant approach relies on engineering the down-conversion process to emit PDC states occupying a single spectral and spatial mode. In the spatial degree of freedom waveguides can be used to restrict the signal and idler beams to the fundamental mode \cite{christ_spatial_2009}. In the spectral degree of freedom, however, a pulsed laser system, appropriately chosen materials and wavelengths have to be applied \cite{grice_spectral_1997, mosley_heralded_2008}.

For PDC processes which are engineered to emit beams into a \textit{single} spatial and spectral mode the generated output state corresponds to a twin-beam squeezed state \cite{barnett_methods_2003}
\begin{align}
    \nonumber
    \ket{\psi}_{PDC} &= \exp\left[r \hat{A}^\dagger \hat{B}^\dagger -  r \hat{A} \hat{B} \right] \ket{0}\\
    &= \sech(r)\sum_{n=0}^{\infty} \tanh^n(r)\ket{n_s, n_i}
    \label{eq:pdc_state}
\end{align}
where we set the phase factor to \(\pi\) as it is unimportant within the scope of this paper. We used capital operators \(\hat{A}\) and \(\hat{B}\) for the signal and idler beam \cite{rohde_spectral_2007} to highlight the pulsed nature of the output state. With this state devoid of multiple spatial and spectral modes the remaining limitations for the heralding of single-photons stem from higher-order photon-number components: Detecting the photons in idler projects the signal into a mixture of photon-number states and, hence, decreases the purity of the heralded state.

\section{Heralding single photons from single-mode PDC sources\label{sec:heralding_single-photons_from_single-mode_PDC_sources}}
Following the discussion of PDC in the previous chapter we now calculate the attainable heralding rates and single-photon fidelities using the state in Eq. \eref{eq:pdc_state} and either binary or photon-number resolving detectors.

The most common method to herald single-photon states from PDC employs binary avalanche photo detection. Depending on its efficiency \(\eta\) it yields a ``Click'' event when photons are measured and a ``NoClick'' event when no photons are detected. Its measurement operators---as a positive operator valued measure (POVM)---are given by \cite{silberhorn_detecting_2007}
\begin{align}
    \nonumber
    \hat{\Pi} _{\mathrm{"No Click"}} &= \sum_{n=0}^{\infty} \left(1- \eta\right)^n \ket{n}\bra{n}\\
    \hat{\Pi}_{\mathrm{"Click"}} &= \sum_{n=0}^{\infty} \left[1-\left(1- \eta\right)^n \right] \ket{n}\bra{n}.
    \label{eq:povm_binary}
\end{align} 
Another approach relies on performing photon-number-resolved measurements in the heralding arm, which are able to enhance the heralding of single-photon states by suppressing the higher photon number components. In past years great advances have been made in photon-number-resolved detection and state-of-the-art detectors feature high detection efficiencies and exhibit an increasing fidelity resolving higher photon numbers. The POVM elements of a general photon-number-resolving detector measuring \(n\) photons are given by  \cite{kok_linear_2007, osullivan_conditional_2008}
\begin{align}
    \hat{\Pi}(n) = \sum_{N=n}^\infty {N \choose n} \left(1-\eta\right)^{N-n} \eta^n \ket{N}\bra{N}
    \label{eq:pnr_povm}
\end{align}
where we assume that each photon has a loss probability of \(\eta\). Individual detection systems may differ from this POVM but all converge to \(\hat{\Pi}(n) = \ket{n}\bra{n}\) for perfect photon-number resolved detection. In the scope of this paper we restrict ourselves to the heralding of single photons, hence, \(n = 1\).

Comparing Eq. \eref{eq:povm_binary} and Eq. \eref{eq:pnr_povm} we notice that both operations have the same structure and are of the form
\begin{equation}
    \hat{\Pi}_{c_n} = \sum_{n=0}^{\infty} c_n \ket{n}\bra{n}
    \label{eq:general_povm}
\end{equation}
where the \(c_n\) coefficients depend on the applied detector and its efficiency \(\eta\). We note that in this formalism dark count events of imperfect detectors could also be included by adapting the \(c_0\) coefficient.

Starting with the single-mode PDC state in Eq. \eref{eq:pdc_state} and the general measurement operator in  Eq. \eref{eq:general_povm}, we calculate the probability of a successful heralding event to be:
\begin{align}
    \nonumber
    p\,(r, c_n) &= _{PDC}\bra{\psi} \hat{\Pi}_{c_n} \ket{\psi}_{PDC}\\
    &= \sech^2(r) \sum_{n=0}^\infty  c_n \tanh^{2n}(r)  
    \label{eq:general_heralding_probability}
\end{align}
and the heralded signal state after a successful detection takes the form
\begin{align}
    \nonumber
    \rho_s(r, c_n) &= \frac{\mathrm{tr}_i \left( \hat{\Pi}_{c_n} {\ket{\psi}_{PDC}}  _{PDC}\bra{\psi} \right)}{_{PDC}\bra{\psi}\hat{\Pi}_{c_n} \ket{\psi}_{PDC}} \\
    &= \frac{\sum_{n=0}^\infty c_n \tanh^{2n}(r) \ket{n_s}\bra{n_s}}{\sum_{n=0}^\infty  c_n \tanh^{2n}(r)}.
    \label{eq:general_heralded_signal_state}
\end{align}
The fidelity of the heralded signal state in Eq. \eref{eq:general_heralded_signal_state} against a pure single-photon is \cite{jozsa_fidelity_1994}:
\begin{align}
    \nonumber
    F(r, c_n) &= \bra{1} \rho_s \ket{1}\\
    &= \frac{c_1 \tanh^2(r)}{\sum_{n=0}^\infty c_n \tanh^{2n}(r) }
    \label{eq:general_heralded_state_fidelity}
\end{align}
Eq. \eref{eq:general_heralding_probability} and Eq. \eref{eq:general_heralded_state_fidelity} form our benchmarks for the state generation: The heralding probability per pulse \(p\,(r, c_n)\) and the fidelity of the generated signal state \(F(r, c_n)\).

\begin{figure}[htb]
    \begin{center}
        \includegraphics[width=\linewidth]{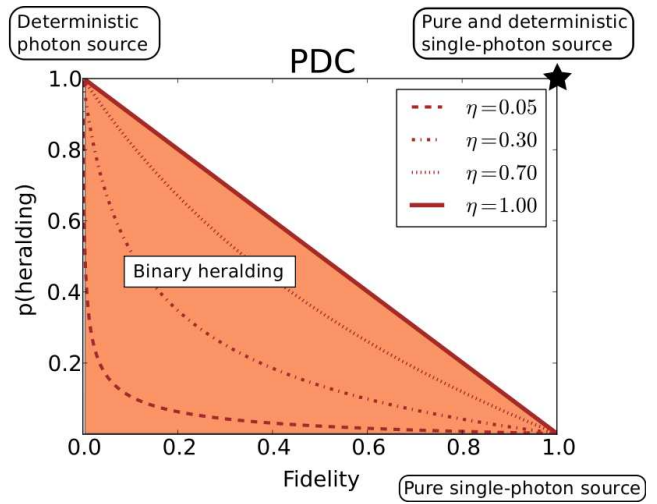}
    \end{center}
    \caption{Heralding probabilities, \(p(heralding)\), and single-photon Fock-state fidelities, \(Fidelity\), of the state created using a single-mode PDC source in conjunction with a binary detector featuring various detection efficiencies, \(\eta\). In this configuration one has to balance either high single-photon fidelities or high state generation rates.}
    \label{fig:binary_photon_heralding}
\end{figure}

In Fig. \ref{fig:binary_photon_heralding} we plotted these benchmarks for a binary detector as given in Eq. \eref{eq:povm_binary} exhibiting various detection efficiencies \(\eta\). The \(x\) axis depicts the achievable fidelities and the \(y\) axis the corresponding heralding probabilities. A source creating perfectly pure single-photon Fock states would appear on the right a source with unit creation probability on the top of the figure. The desired pure deterministic single-photon source resides in the upper right corner of the graphic.
The shaded region in Fig. \ref{fig:binary_photon_heralding} depicts the general area available using PDC in conjunction with binary detectors and presents an inherent trade-off between signal creation rate and fidelity of the heralded state. Even with perfect detectors \(\eta = 1\), either the PDC process only emits photon pairs (\(r \le 0.1\)), which yields near unit fidelities but low heralding rates, or one can choose PDC states with higher-order photon-number components leading to heralding probabilities approaching unity (\(r \ge 2\)) yet at the cost of low fidelities due to the occurring mixing in photon number. 

\begin{figure}[htb]
    \begin{center}
        \includegraphics[width=\linewidth]{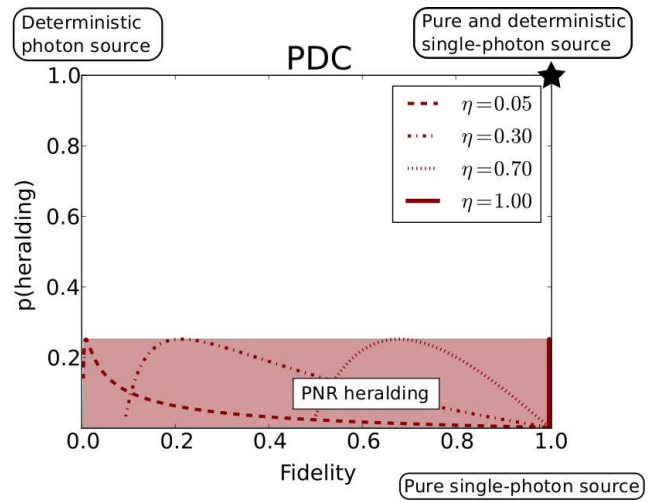}
    \end{center}
    \caption{Heralding probabilities, \(p(heralding)\), and single-photon Fock-state fidelities, \(Fidelity\), of the state created using a single-mode PDC source in conjunction with a photon-number resolving detector featuring various detection efficiencies, \(\eta\). These detectors suppress higher-order photon numbers and, hence, enable high fidelities in conjunction with heralding rates ranging up to 25\%.}
    \label{fig:pnr_photon_heralding}
\end{figure}

In Fig. \ref{fig:pnr_photon_heralding} we plotted the heralding probability \(p(r, c_n)\) and the state fidelity \(F(r, c_n)\) using a photon-number-resolving detector as defined in Eq. \eref{eq:pnr_povm} for various detection efficiencies \(\eta\). It is evident that photon-number-resolving detectors are superior to binary detectors. They enable unit fidelities in conjunction with heralding rates up to \(25\%\) only constrained by the thermal photon-number distribution emitted by the down-conversion process (\(p_{th}^{(max)}(1)=25\%\)). In the case of perfect detection \(\eta = 1\), this figure gives the fundamental limit of PDC sources. Creating perfectly pure single-photon Fock states the maximum achievable heralding rate is \(25\%\). The corresponding PDC source features an amplitude of \(r=0.88\), corresponding to a squeezing value of 7.64 dB and a mean-photon number of \(\langle n_{ph} \rangle = 1\). 

\section{Heralding single photons from multimode PDC sources\label{sec:heralding_single-photons_from_multimode_PDC_sources}}
We now turn our attention to the impact of spectral multimode effects on the heralding rates and single-photon fidelities. While it is relatively straightforward to get rid of spatial multimode effects in the PDC state emission it is not trivial to construct a source which only emits into a single spectral mode \cite{mosley_heralded_2008, eckstein_highly_2011}. Hence, we extend our analysis and investigate spectrally multimode PDC as a source of single-photon states in order to evaluate to which degree multimode spectral components can be tolerated.

Including multiple spectral modes the PDC state in Eq. \eref{eq:pdc_state} takes the form \cite{christ_probing_2011}:
\begin{align}
    \nonumber
    \ket{\psi}_{PDC} =&  \exp\left[-\frac{\imath}{\hbar} \left(B \iint \mathrm d\omega_s d \omega_i f(\omega_s, \omega_i) \right. \right. \\ 
    \nonumber
    &\times \left. \left .\hat{a}^{(s)\dagger}(\omega_s) \hat{a}^{(i)\dagger}(\omega_i) + h.c. \right)\right] \ket{0} \\
    =& \bigotimes_k \exp\left[r_k \hat{A}^\dagger_k \hat{B}^\dagger_k - r_k \hat{A}_k \hat{B}_k \right] \ket{0} \\
    =& \bigotimes_k \sech(r_k)\sum_{n=0}^{\infty} \tanh^n(r_k)\ket{n_k^{(s)}, n_k^{(i)}}
    \label{eq:pdc_state_multimode_frequency}
\end{align}
In this case, not a single twin-beam squeezed state, as depicted in Eq. \eref{eq:pdc_state}, is generated, but a multitude of twin-beam squeezers with amplitudes \(r_k\) in broadband frequency modes \(\hat{A}_k\) and \(\hat{B}_k\) \cite{rohde_spectral_2007} are emitted. For common PDC sources the squeezer distribution \(r_k\) follows an exponential decay \cite{uren_photon_2003} and is defined by
\begin{align}
    \nonumber
    &r_k = B \lambda_k \\
    &\lambda_k =\sqrt{1-\mu}\, \mu^k \qquad 0 \le \mu \le 1
    \label{eq:rk_distribtion}
\end{align}
where \(B\) is optical gain depending on the applied nonlinearity, on the pump power in the PDC process and \(\lambda_k\) corresponds to the normalized mode distribution. The \textit{effective} number of optical modes in the state is quantified by the parameter \(K = 1 / \sum_k \lambda_k^4\) \cite{eberly_schmidt_2006}. 

The properties of the PDC state in Eq. \eref{eq:pdc_state_multimode_frequency} become clearer if we sort the terms according to their photon-number components,
\begin{align}
    \nonumber
    \ket{\psi}_{PDC} =& A \ket{0} + A \sum_k \tanh(r_k) \ket{1_k; 1_k} \\
    \nonumber
    &+ A \sum_{k \le k'} \tanh(r_k) \tanh(r_{k'}) \ket{1_k, 1_{k'}; 1_k, 1_{k'}} \\
    &+ \dots \, ,
    \label{eq:mm_pdc_state_expanded}
\end{align}
where \(A = \prod_l \sech(r_l) \), \(\ket{0} = \bigotimes_k \ket{0_k}\) and \(\varphi_k = \pi\). According to Eq. \eref{eq:mm_pdc_state_expanded} the PDC state now consists of multiple photon-pair components emitted into an array of spectral modes \(k\).

Given a multimode PDC state as defined in Eqs. \eref{eq:pdc_state_multimode_frequency} and \eref{eq:mm_pdc_state_expanded}, we calculate the heralding rates and fidelities similar to Sec. \ref{sec:heralding_single-photons_from_single-mode_PDC_sources}. In order to perform this calculation we extend the measurement operators given in Eq. \eref{eq:povm_binary}, Eq. \eref{eq:pnr_povm} and Eq. \eref{eq:general_povm} to the multimode regime:
\begin{align}
    \nonumber
    \hat{\Pi}_{c_n} =& c_0 \ket{0}\bra{0} + c_1 \sum_k \ket{1_k}\bra{1_k} \\
    &+ c_2 \sum_{k \le k'} \ket{1_k, 1_{k'}}\bra{1_k, 1_{k'}} + \dots
    \label{eq:povm_general_multimode}
\end{align}
where the \(c_n\) terms are identical to the single-mode case as we assume that the detector cannot distinguish different frequencies due to limited time resolution. Using Eq. \eref{eq:mm_pdc_state_expanded} and Eq. \eref{eq:povm_general_multimode} we obtain a multimode heralding probability of
\begin{align}
    \nonumber
    p(r_k, c_n) =& c_0 A^2 + c_1 A^2 \sum_k \tanh^2(r_k) \\
    +& c_2 A^2 \sum_{k \le k'} \tanh^2(r_k) \tanh^2(r_{k'}) + \dots
    \label{eq:general_heralding_propabiltiy_multimode}
\end{align}
and the heralded signal state takes on the form,
\begin{align}
    \nonumber
    \rho_s =& \frac{1}{N} c_0 \ket{0}\bra{0} + c_1 \sum_k \tanh^2(r_k) \ket{1_k}\bra{1_k} \\
    \nonumber
    &+ c_2 \sum_{k \le k'} \tanh^2(r_k) \tanh^2(r_{k'}) \ket{1_k, 1_{k'}}\bra{1_k, 1_{k'}} \\
    &+ \dots
    \label{eq_heralded_signal_state_multimode}
\end{align}
with the normalization constant \(N\) defined as
\begin{align}
    \nonumber
    N =&c_0 + c_1 \sum_k \tanh^2(r_k) \\
    & + c_2 \sum_{k \le k'} \tanh^2(r_k) \tanh^2(r_{k'}) + \dots \, .
    \label{eq:general_heralded_state_fidelity_multiple_normalisation}
\end{align}
The corresponding fidelity of the heralded photon state against a single-photon Fock state evaluates to
\begin{align}
    F(r_k, c_n) = \frac{1}{N} c_1 \tanh^2(r_0).
    \label{eq:general_heralded_state_fidelity_multimode}
\end{align}
Eq. \eref{eq:general_heralding_propabiltiy_multimode} and \eref{eq:general_heralded_state_fidelity_multimode} enable us to benchmark multimode PDC processes as a source of heralded single-photon states via the heralding probability \(p(r_k, c_n)\) and the fidelity \(F(r_k, c_n)\) of a heralded state including \textit{both} the spectral and photon-number degree of freedom. Note that the performance of spectrally filtered PDC states will lie below a spectrally single-mode source.

\begin{figure}[htb]
    \begin{center}
        \includegraphics[width=\linewidth]{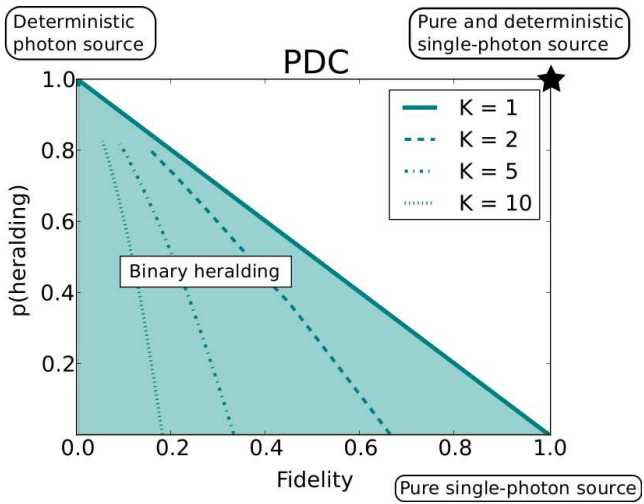}
    \end{center}
    \caption{Heralding probabilities, \(p(heralding)\), and single-photon Fock-state fidelities, \(Fidelity\), of the state created using various multimode PDC sources in conjunction with a binary detector featuring unit detection efficiency \(\eta = 1\). Multiple frequency modes \(K = 2, 5, 10\) severely limit the achievable maximum fidelities}.
    \label{fig:binary_photon_heralding_multimode}
\end{figure}
We visualized the obtained rates and fidelities using a binary detector with efficiency \(\eta = 1\) in Fig. \ref{fig:binary_photon_heralding_multimode}. In this figure we use four exemplary PDC states with rising \textit{effective} mode numbers \(K = 1, 2, 5, 10 \), where \(K=1\) corresponds to the single-mode case discussed in Sec. \ref{sec:heralding_single-photons_from_single-mode_PDC_sources}. Fig. \ref{fig:binary_photon_heralding_multimode} shows that the mixing in frequency diminishes the maximal attainable fidelities over the whole range of heralding rates (\(K=2,5,\) and 10 plotted up to \(B = 1.36\)). 

\begin{figure}[htb]
    \begin{center}
        \includegraphics[width=\linewidth]{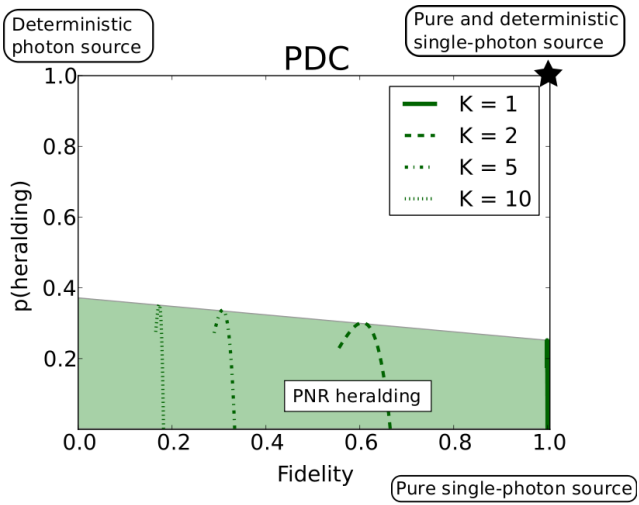}
    \end{center}
    \caption{Heralding probabilities, \(p(heralding)\), and single-photon Fock-state fidelities, \(Fidelity\), of the state created using various multimode PDC sources in conjunction with a photon-number-resolving detector featuring unit detection efficiency \(\eta = 1\). Again, multiple frequency modes \(K = 2, 5, 10\) severely limit the achievable maximum fidelities}
    \label{fig:pnr_photon_heralding_multimode}
\end{figure}

This mixing in frequency modes also negatively affects photon-number resolved detection. Fig. \ref{fig:pnr_photon_heralding_multimode} depicts the heralding of single photons from a multimode PDC state using a photon-number-resolving detector with efficiency \(\eta=1\). Again the maximum achievable fidelities are constrained by the number of optical modes in the PDC.

In total multimode spectral effects ultimately limit the  achievable heralded single-photon fidelities. This issue consequently must be addressed by generating the PDC state in a single spectral mode as discussed in Sec. \ref{sec:heralding_single-photons_from_single-mode_PDC_sources}. Alternatively, filtering the idler beam can be applied to create single photons in a single spectral mode, yet at the expense of severe losses in photon number.

\section{Deterministic pure-single-photon \\ generation with switched PDC sources }
Our previous calculations showed that it is impossible to build a pure deterministic single-photon source using a single PDC process. However, it has been noted that multiple PDC sources in a switched setup may be able to create a source approximating deterministic behavior \cite{migdall_tailoring_2002, pittman_single_2002, ma_experimental_2011, jennewein_single-photon_2011}. This approach employs multiple PDC single-photon sources: When one signals the successful heralding of a single-photon state the photon is routed to the output. Given a photon heralding probability of \(\nu\) and lossless routing the overall heralding probability in a switched setup---as a function of the number of applied PDC sources \(n\)---is:
\begin{align}
    p(\mathrm{"switched"}) = 1 - \left(1- \nu\right)^n
    \label{eq:switching_propability}
\end{align}

\begin{figure}[htb]
    \begin{center}
        \includegraphics[width=\linewidth]{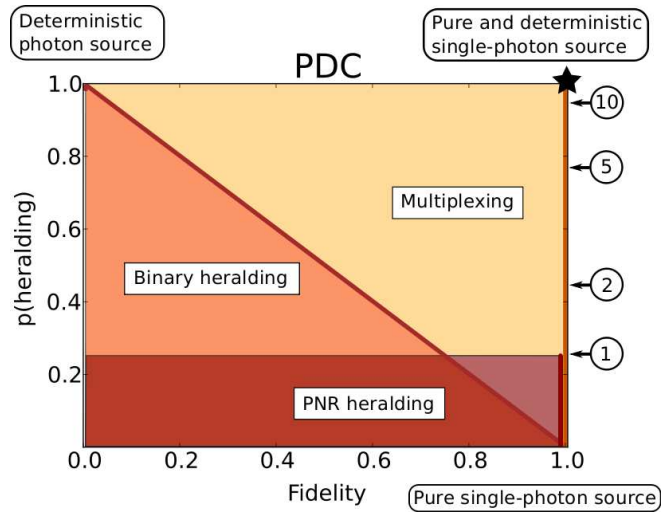}
    \end{center}
    \caption{Heralding probabilities, \(p(heralding)\), and fidelities, \(Fidelity\), accessible using a single-mode PDC source in conjunction with a binary detector (orange-shaded region), a photon-number resolving detector (red-shaded region), and multiplexing (yellow-shaded region). For an optimal source and a perfect photon-number-resolving detector with heralding probability of \(\nu = 25\%\) 17 PDC sources are required to obtain a deterministic single-photon source (\(>99\%\) emission probability). The arrows point out the achievable heralding rates using a multiplexed setup of 1, 2, 5, and 10 single photon sources.}
    \label{fig:overview}
\end{figure}
Fig. \ref{fig:overview} presents the impacts of multiplexing on the rates and fidelities and  summarizes our results. A single-mode PDC source in conjunction with binary detectors suffers from an inherent trade-off between high heralding rates and high fidelities (orange-shaded region). Photon-number-resolving detectors solve this issue and enable heralding efficiencies up to 25\% and unit fidelities (red-shaded region). Multiplexing these single PDC setups enables access to sources featuring high heralding rates in conjunction with unit fidelities (yellow-shaded region). The achievable rates for the multiplexing of 1, 2, 5, and 10 PDC sources are displayed in Fig. \ref{fig:overview}.

The overhead in the number of PDC sources is, of course, quite significant. Hence, the most practicable route to create deterministic pure single-photon Fock states using PDC is first to move from binary to photon-number-resolved detection which enables unit fidelities and significant heralding rates for a single source. Multiplexing these setups gives access to the desired pure deterministic behavior. Given optimal PDC sources with perfect photon-number resolved-detection (\(\nu = 25\%\)) 17 PDC setups are required to approximate a deterministic pure single-photon source (\(>99\%\)).

\section{Conclusion}
In conclusion we determined the prospects for PDC to serve as a \textit{pure} deterministic single-photon source. We investigated the effects of the spectral \textit{and} the photon-number degree of freedom on heralding pure single-photon states from PDC. Our findings show that the spectral degree of freedom limits the achievable fidelities of the heralded signal states and hence spectral effects have to be negated by engineering of the PDC process to occupy a single spectral mode. 

For a PDC state free of multiple spectral modes the remaining limitations stem from the higher-order photon components and the applied detectors. Binary detectors feature an inherent trade-off between high heralding probability and near unit state fidelity, whereas photon-number-resolving detectors are able to herald \textit{pure} single-photon Fock states with a probability of up to \(25\%\), given unit detection efficiencies and an optimal PDC state with a twin-beam squeezing of 7.64 dB (\(\langle n_{ph} \rangle  = 1 \)). This forms the fundamental limit on heralding pure single-photon states using PDC. Applying a switched PDC setup to increase the heralding rate 17 individual sources are, hence, required to approximate a pure deterministic single-photon source (\(>99\%\) emission probability).   

\section*{Acknowledgements}
The research leading to these results has received funding from the European Community’s Seventh Framework Programme FP7/2007-2013 under the grant agreement Q-Essence 248095. The authors thank Kaisa Laiho, Malte Avenhaus, Helge Rütz, and Benjamin Brecht for useful discussions and helpful comments.

\bibliography{limits_on_pdc.bib}

\end{document}